\documentclass[12pt]{article}
\usepackage[utf8]{inputenc}
\usepackage[normalem]{ulem} 
\usepackage{latexsym,color}
\usepackage{cite,epsfig,amssymb,euscript,slashed}
\usepackage{amsmath}
\usepackage{array,calc,epsfig}
\usepackage[colorlinks,allcolors=blue]{hyperref}
\usepackage{bbm}
\usepackage{fancybox}

\def\be{\begin{equation}}
\def\ee{\end{equation}}
\def\bseq{\begin{subequations}}
\def\eseq{\end{subequations}}

\def\bea{\begin{eqnarray}}
\def\eea{\end{eqnarray}}

\def\bseq{\begin{subequations}}
\def\eseq{\end{subequations}}

\numberwithin{equation}{section} 
\usepackage[]{graphicx}

\def\ii           {{\rm i}}

\def\tr           {\mathop{\rm Tr}}

\def\sqr#1#2{{\vcenter{\vbox{\hrule height.#2pt
 \hbox{\vrule width.#2pt height#1pt \kern#1pt \vrule width.#2pt}\hrule
 height.#2pt}}}}

\def\slashchar#1{\setbox0=\hbox{$#1$}           
\dimen0=\wd0                                 
\setbox1=\hbox{/} \dimen1=\wd1               
\ifdim\dimen0>\dimen1                        
\rlap{\hbox to \dimen0{\hfil/\hfil}}      
#1                                        
\else                                        
\rlap{\hbox to \dimen1{\hfil$#1$\hfil}}   
/                                         
\fi}

\begin{document}
\font\cmss=cmss10 \font\cmsss=cmss10 at 7pt

\title{\vspace{-2.0cm}
Can Chern-Simons or Rarita-Schwinger be a Volkov-Akulov Goldstone?
\\[0.5cm] }

\author{Sukruti Bansal\footnote{e-mail: {\tt  bansal.sukruti@gmail.com}} \,\,and Dmitri Sorokin\footnote{e-mail: {\tt  dmitri.sorokin@pd.infn.it }}}

\date{}

\maketitle

\vspace{-1.5cm}

\begin{center}

\vspace{0.5cm}
\textit{\small  Dipartimento di Fisica e Astronomia ``Galileo Galilei",  Universit\`a degli Studi di Padova \\
\& I.N.F.N. Sezione di Padova, Via F. Marzolo 8, 35131 Padova, Italy}
\end{center}

\vspace{5pt}

\abstract{We study three-dimensional non-linear models of vector and vector-spinor Goldstone fields associated with the spontaneous breaking of certain higher-spin counterparts of supersymmetry whose Lagrangians are of a Volkov-Akulov type. Goldstone fields in these models transform non-linearly under the spontaneously broken rigid symmetries. We find that the leading term in the action of the vector Goldstone model is the Abelian Chern-Simons action whose gauge symmetry is broken by a quartic term. As a result, the model has a propagating degree of freedom which, in a decoupling limit, is a quartic Galileon scalar field. The vector-spinor goldstino model turns out to be a non-linear generalization of the three-dimensional Rarita-Schwinger action. In contrast to the vector Goldstone case, this non-linear model retains the gauge symmetry of the Rarita-Schwinger action and eventually reduces to the latter by a non-linear field redefinition. We thus find that the free Rarita-Schwinger action is invariant under a hidden rigid supersymmetry generated by fermionic vector-spinor operators and acting non-linearly on the Rarita-Schwinger goldstino.}


\thispagestyle{empty}



\setcounter{footnote}{0}


%
\newpage

\section{Introduction}
In 1975 Hietarinta \cite{Hietarinta:1975fu} constructed (graded) Lie algebras which are a higher-spin generalization of the conventional Poincar\'e superalgebras. Instead of spinorial supersymmetry generators  associated with spin-$\frac 12$, these algebras include (spinor-)tensor generators associated with (half-)integer higher-spin representations of the Lorentz group. As in the supersymmetry case, (anti-)commutators of these generators close on the generator of space-time translations. The D-dimensional Hietarinta algebras have the following generic structure
\bea\label{Ha}
\{Q^{a_1...a_n}_\alpha,Q^{b_1...b_m}_\beta\}=f^{a_1...a_n,b_1...b_m,c}_{\alpha\beta}P_c\,,\nonumber\\
{}[S^{a_1...a_p},S^{b_1...b_q}]=f^{a_1...a_n,b_1...b_m,c}P_c\,,\nonumber\\
{}[Q,P]=0\,, \qquad [S,P]=0\,,\qquad [Q,S]=0\,,
\eea
where $a,b,c=0,1...,D-1$ are vector indices, $\alpha,\beta$ are spinor indices, $Q^{a_1...a_n}_\alpha$ are fermionic tensor-spinor generators, $S^{a_1...a_p}$ are bosonic tensor generators and $P_c$ is the translation generator. The generators transform under certain representations of the Lorentz group $S=SO(1,D-1)$. The structure constants $f^{a_1...a_n,b_1...b_m,c}_{\alpha\beta}$ and $f^{a_1...a_n,b_1...b_m,c}$ are $SO(1,D-1)$ invariant
and constructed with the use of the Minkowski metric, Levi-Civita tensor and gamma-matrices.

The algebras \eqref{Ha} are finite-dimensional higher-spin algebras. This distinguishes them from the more familiar infinite-dimensional higher-spin algebras in which the (anti)-commutators of higher-spin generators close on generators carrying yet higher spins.

For building models with spontaneously broken symmetries of this kind Hietarinta used the Volkov-Akulov construction of Lagrangians with non-linearly realized supersymmetry \cite{Volkov:1972jx,Volkov:1973ix}. The case of a $D=4$ spin-$\frac 32$ superalgebra and its non-linear realizations was independently considered in \cite{Baaklini:1977ka} (see also \cite{Pilot:1988st,Pilot:1988gy}) and further exploited e.g. in \cite{Shima:2000fs} and references therein. In four-dimensional space-time, consistency issues of a gravitational coupling of a massless spin-$\frac 52$ field, which might be regarded as a gauge field of the local spin-$\frac 32$ supersymmetry, were studied  were studied in \cite{Berends:1979wu,Berends:1979rv,Berends:1979kg,Aragone:1979hx}.

In three space-time dimensions, however,  Aragone and Deser \cite{Aragone:1983sz} succeeded in constructing a consistent  `hypergravity' model which is invariant under local symmetry transformations associated with a spin-$ (n+\frac 12)$ superalgebra ($n=0,1,...$) and describes interacting non-propagating graviton and a spin-$ (n+\frac 32)$ gauge field\footnote{Strictly speaking, as is well known, in $D=3$ massless representations of the Poincar\'e group are spin-less. However, as is often adopted in higher-spin literature for any space-time dimension, we loosely call symmetric tensor fields $A_{a_1\ldots a_s }$ of rank $s$  as integer spin-$s$ fields and symmetric-tensor spinor fields $\Psi^\alpha_{a_1\ldots a_s}$ as half-integer spin $s+\frac 12$ fields.}. Much more recently this model was extended to an $AdS_3$ background including an additional spin-4 field by Zinoviev \cite{Zinoviev:2014sza} who also constructed its higher-spin generalizations. Different aspects of higher-spin superalgebras of this kind in $D\geq 3$ and associated models were also considered in \cite{Bunster:2014fca,Fuentealba:2015jma,Fuentealba:2015wza,Henneaux:2015tar}. It may be of interest to study the effects of spontaneous symmetry breaking in these models, which is one of the motivations of this paper.

More general motivation is related to the fact that, as is well-known, the construction of interacting higher-spin theories in space-time dimensions higher than three is a highly non-trivial problem\footnote{For a review of various aspects of higher-spin field theory and references see e.g. \cite{Vasiliev:1995dn,Buchbinder:1995uq,Vasiliev:1999ba,Vasiliev:2001ur,Bekaert:2003uc,Sorokin:2004ie,Bouatta:2004kk,Bandos:2005rr,Bekaert:2005vh,Francia:2006hp,Fotopoulos:2008ka,Campoleoni:2009je,Taronna:2010qq,Bekaert:2010hw,Sezgin:2012ag,Gaberdiel:2012uj,Giombi:2012ms,Joung:2012fv,Sagnotti:2013bha,Didenko:2014dwa,Rahman:2015pzl,Arias:2016ajh,Sleight:2016hyl,Sleight:2017krf,Sorokin:2017irs,Rahman:2017cxk}.}. This issue also regards models based on the higher-spin algebras of \cite{Hietarinta:1975fu}. In \cite{Shima:1989nb} it was shown (for the spin-$\frac 32$ case in $D=4$) that these algebras do not have non-trivial linear unitary representations. Yet, one may still ask the question whether the higher-spin Goldstone field constructions based on the non-linear realizations of these algebras produce physically consistent interacting models. A priori, such a possibility is not excluded, since non-linearly realized symmetry may act only on positive-norm states while the negative-norm states of corresponding linear multiplets are cut off.

To the best of our knowledge the physical properties and the consistency of the Goldstone models associated with this type of higher-spin algebras have not yet been considered in the literature (even for the simplest cases of spin-1 and spin-$\frac 32$), and this is the purpose of our paper.
We will study this problem in three-dimensional space-time for Goldstone fields of spin-1 and spin-$\frac 32$. As we will see, these simplest models already exhibit particular, interesting features. The leading term in the action of the spin-1 Goldstone model is the Abelian Chern-Simons Lagrangian whose gauge symmetry is broken by a quartic term. As a result, the model has a propagating degree of freedom which, in a decoupling limit, is a quartic Galileon scalar field. The Hamiltonian of this model is not bounded from below signalling the presence of instabilities. At the same time, somewhat surprisingly, the vector-spinor Goldstino model, which is a non-linear generalization of the three-dimensional Rarita-Schwinger Lagrangian, does possess a non-linearly extended local symmetry of the Rarita-Schwinger Lagrangian. Hence, it does not have propagating degrees of freedom. Moreover, as we will see, the non-linear spin-$\frac32$ goldstino action reduces to the free Rarita-Schwinger action by a non-linear field redefinition. We thus find that the free Rarita-Schwinger action is invariant under a hidden non-linearly realized rigid supersymmetry generated by fermionic vector-spinor operators and that the Rarita-Schwinger field is the goldstino field associated with the spontaneous breaking of this symmetry.

The paper is organized as follows. In Section \ref{12} we will review the Volkov-Akulov construction of the Lagrangian for a goldstino field associated with the spontaneous breaking of the conventional $N=1$ supersymmetry, whose Poincar\'e superalgebra in $D=3$ has the following form:
\bea \label{poincare}
[M_{ab}\,,M_{cd}]&=&\ii(\eta_{bc}\,M_{ad}-\eta_{ac}\,M_{bd}-\eta_{bd}\,M_{ac}+\eta_{ad}\,M_{bc})\,,\nonumber \\
{}[M_{ab}\,,P_c]&=&\ii(\eta_{bc}\,P_a-\eta_{ac}\,P_b)\,, \nonumber \\
{}[P_a,P_b]&=&0\,,
\eea
\bea\label{susy12}
{}[M_{ab}\,,Q_\alpha]&=&-\,\frac{\ii}{2}\,(\Gamma_{ab})_\alpha{}^\beta \,Q_\beta \,,\nonumber \\
\{Q_\alpha\,,Q_\beta\} &=&\,2\,(\Gamma^a C^{-1})_{\alpha \beta}\,P_a\,, \nonumber \\
{}[Q_\alpha\,,P_a]&=&0\,,
\eea
where $M_{ab}$ ($a,b=0,1,2$) is the generator of the Lorentz group $SO(1,2)$, $P_a$ is the translation generator and $Q_\alpha$ ($\alpha=1,2$) is the Majorana spinor generator of the supersymmetry transformations. We use the ``mostly plus" convention for the Minkowski metric and the real Majorana representation for the gamma-matrices (see the Appendix for more details).

As an instructive exercise, we will explicitly check that the higher-order terms in the Volkov-Akulov Lagrangian give a positive-definite contribution to the Hamiltonian, thus demonstrating the fact that the non-linear Volkov-Akulov goldstino model does not have ghosts.

In Section \ref{1} we will apply the Volkov-Akulov procedure to the construction of a model describing a spin-1 goldstone field associated with spontaneous breaking of a spin-1 counterpart of the $N=1$ superalgebra \eqref{susy12}. Spin-1 algebra is generated by Poincar\'e generators \eqref{poincare} and a bosonic vector operator $S^a$ satisfying the following commutation relations:
\be\label{vector}
{}[M^{ab}\,,S^c]=\ii(\eta^{bc}\,S^a-\eta^{ac}\,S^b)\,,
\ee
\be\label{spin1}
{}[S^a,S^b]=2\ii\,\varepsilon^{abc}P_c\,,\qquad {}[S^a,P_b]=0\,.
\ee
Note in passing, that the algebra \eqref{spin1} can be regarded as an Inonu-Wigner contraction of the $so(2,2)$-algebra.

The Goldstone field associated with $S^a$ is a vector field $A_a(x)$. As we will see, the Volkov-Akulov-type model for this field is described by an action whose quadratic part is the standard Abelian Chern-Simons action. The latter is invariant under the gauge transformations $A_a(x) \rightarrow A_a(x)+\partial_a\lambda (x)$ which make the Chern-Simons field non-dynamical, as is of course well known. We will study whether the complete non-linear action for the Goldstone vector field still possesses (a non-linear generalization of) this gauge symmetry and find that this is not the case. To this end, we will carry out the Dirac analysis of constrained Hamiltonian systems (see e.g. \cite{Dirac-lectures,HT-book}). We will show that for generic classical field configurations the non-linear model under consideration does not have first-class constraints associated with local gauge symmetries, but only second-class ones. As a result, it contains one St\"uckelberg-like scalar propagating degree of freedom whose Lagrangian, in a decoupling limit, turns out to be the same as the quartic Galileon Lagrangian \cite{Nicolis:2008in} but with a missing quadratic kinetic term. The Hamiltonian of this model is unbounded from below. Hence fluctuations around certain zero-energy backgrounds may have a negative energy and lead to instabilities. These instabilities are not of the (higher-derivative) Ostrogradski type, since the higher-order Galileon Lagrangians are quadratic in time derivatives. Note, in passing, that due to their peculiar properties, Galileon models have been intensively studied in the theories of modified gravity and cosmology. For a review see e.g. \cite{Hinterbichler:2011tt,deRham:2014zqa,Schmidt-May:2015vnx,Amendola:2016saw} and the references therein.

In Section \ref{s32} we will consider the case of a spin-$\frac 32$ Goldstone field model associated with the spin-$\frac 32$ superalgebra \cite{Hietarinta:1975fu,Baaklini:1977ka,Pilot:1988st} whose most general form in $D=3$ is
\be\label{32}
{}[M^{ab}\,,Q_\alpha^c]=\ii(\eta^{bc}\,Q_\alpha^a-\eta^{ac}\,Q_\alpha^b)-\frac{\ii}{2}\,(\Gamma^{ab})_\alpha{}^\beta \,Q^c_\beta\,,
\ee
\be\label{spin32}
{}\{Q^a_\alpha,Q^b_\beta\}=2\,{\tt a}\,C_{\alpha\beta}\,\varepsilon^{abc}\,P_c+{\tt b}\,\Gamma^{(a}_{\alpha\beta}P^{b)}+{\tt c}\,\eta^{ab}\,\Gamma^c_{\alpha\beta}\,P_c\,,\qquad {}[Q^a_\alpha,P_b]=0\,,
\ee
where $\tt a$, $\tt b$ and $\tt c$ are arbitrary real parameters. One of these parameters can always be set to a given number by re-scaling the fermionic generators $Q^a_\alpha$ or the momentum $P_a$.

Note that, in general, $Q^a_\alpha$ is transformed under a reducible representation of the Lorentz group which splits into the irreducible parts as follows
\be\label{split}
Q^a_\alpha=\hat Q^a_\alpha+\frac 13\, (\Gamma^aQ)_\alpha\,,
\ee
where $Q_\alpha$ is a Majorana-spinor generator and $\hat Q^a_\alpha$ is gamma-traceless ($\Gamma_a\hat Q^a=0$).

Depending on the choice of the parameters ${\tt a},{\tt b}$ and ${\tt c}$, the superalgebra \eqref{spin32} can be reduced to simpler superalgebras. Three specific cases are the following ones.

When ${\tt a}=-\frac 5{12}$, ${\tt b}=\frac 13$ and ${\tt c}=-\frac 23$, the only non-trivial anti-commutator in \eqref{spin32} is between the gamma-traceless $\hat Q^a_\alpha$, while the spin-$\frac 12$ generators $Q_\alpha$ anti-commute with themselves and with $\hat Q^a_\alpha$. This superalgebra was exploited in \cite{Fuentealba:2015jma}.

If instead, ${\tt b}=4\,{\tt a}$ and ${\tt c}=-\,2\,{\tt a}$, only the spin-$\frac 12$ generators $Q_\alpha$ have a non-trivial commutator, as in \eqref{susy12}, while the gamma-traceless generators $\hat Q^a_\alpha$ anti-commute with themselves and with $Q_\alpha$ and hence decouple. Therefore, in this case, the superalgebra \eqref{spin32} reduces to the conventional $N=1$ superalgebra.

The third case is when ${\tt b}={\tt c}=0$ and e.g. ${\tt a}=1$. Then the algebra \eqref{spin32} reduces to
\be\label{spin321}
{}\{Q^a_\alpha,Q^b_\beta\}=2\,C_{\alpha\beta}\,\varepsilon^{abc}P_c\,,\qquad {}[Q^a_\alpha,P_b]=0\,.
\ee
In this paper we will consider the Volkov-Akulov-like model associated with the spin-$\frac 32$ superalgebra of the type \eqref{spin321},
since the quadratic part of its non-linear Lagrangian coincides with the Rarita-Schwinger (or Chern-Simons-like) Lagrangian for a massless spinor-vector field $\chi^a_\alpha$. The gamma-traceless case can be associated with the gauge-fixed Rarita-Schwinger action in which $\Gamma_a\chi^a=0$, while for other (inequivalent) choices of parameters (except those corresponding to the conventional supersymmetry), the spin-$\frac 32$ superalgebra does not seem to produce physically consistent models even in the free (quadratic) approximation because of the absence of gauge symmetry.

We will show that, in contrast to the spin-1 case,  higher-order contributions to the spin-$\frac 32$ goldstino action do not break the gauge symmetry of its quadratic Rarita-Schwinger part but only require a non-linear modification of the gauge variation of the spin-$\frac 32$ field. Moreover, the non-linear action reduces to the free Rarita-Schwinger action by an invertible non-linear field redefinition, which means that the Rarita-Schwinger action itself is non-manifestly invariant under the non-linearly realized spin-$\frac 32$ supersymmetry \eqref{spin321}.

In the Conclusion we will briefly discuss possible extensions of our results. In particular, we will present an action for a 3d gravity model of two spin-2 gauge fields interacting via a Lorentz spin connection, which is invariant under the local symmetries generated by the algebra \eqref{vector} and \eqref{spin1}.

\section{Volkov-Akulov model of the spin-1/2 goldstino}\label{12}
The Volkov-Akulov construction \cite{Volkov:1972jx,Volkov:1973ix}  of the action for a real Majorana-spinor goldstino $\chi^\alpha(x)$ associated with the spontaneous breaking of supersymmetry \eqref{susy12} uses, as a building block, a one-form,\footnote{For a recent review of the different aspects and realizations of the Volkov-Akulov model and its coupling to supergravity, see \cite{Bandos:2015xnf,Bandos:2016xyu} and the references therein.}$^,$\footnote{
As a shorthand notation, in what follows, we define the contraction of the spinors with a single gamma-matrix as $\chi\,\Gamma^a\,\psi\equiv \chi^\alpha\,\Gamma_{\alpha\beta}^a\,\psi^\beta=-\,\chi^\alpha\,\Gamma^{a\,\beta}_{\alpha}\,\psi_\beta$. For other rules regarding the handling of the spinor indices see the Appendix.}
\be\label{VA12}
E^a=dx^a+\ii f^{-2}\,\chi^\alpha(x)\,\Gamma^a_{\alpha \beta}\, d\chi^\beta(x)=dx^b(\delta^a_b+\ii f^{-2}\,\chi\,\Gamma^a\, \partial_b\chi)\equiv dx^bE_b^a\,,
\ee
which is invariant under the following supersymmetry variations of $x^a$ and $\chi^\alpha(x)$ generated by the algebra \eqref{susy12}
\be\label{susyxpsi12}
x'^a{}=x^a-\ii\, f^{-2}\,\epsilon\,\Gamma^a\,\chi\,, \qquad \chi'^{\alpha}(x')=\chi^\alpha(x)+\epsilon^\alpha\,,
\ee
where $\epsilon^\alpha$ is a constant spinor parameter, $f$ is a supersymmetry breaking parameter of mass-dimension $m^{\frac 32}$ and $\chi^\alpha$ has the $D=3$ canonical dimension of $m$. The infinitesimal transformation of the form of the goldstino field $\chi^\alpha(x)$,
\begin{align}\label{vary12}
\delta \chi^\alpha(x)=\epsilon^\alpha+\ii\,f^{-2}\, \big(\epsilon\,\Gamma^a\,\chi(x)\big)\,   \partial_a \chi^\alpha(x)\,,
\end{align}
shows that it transforms non-linearly under supersymmetry. The commutator of two variations \eqref{vary12} closes on the translations off the mass shell, i.e. without the use of the equations of motion.
\begin{align}
[\delta_2\,,\delta_1]\,\chi^\alpha=2\,\ii\, f^{-2}\,(\epsilon_1\,\Gamma^a\,\epsilon_2)\,\partial_a\chi^\alpha\,.
\end{align}

The supersymmetry invariant Volkov-Akulov action in $D=3$ is
\be\label{generic}
S=\frac{f^2}6\int\,E^a\wedge E^b\wedge E^c\,\varepsilon_{abc}=-f^2\int\,d^3x\,\det E^a_b\,,
\ee
or explicitly
\be\label{VAa12}
S_{1/2}=  \int\,d^3x\left(-f^2-\ii\,\chi\,\Gamma^a\, \partial_a\chi+\frac{f^{-2}}{2}\,\varepsilon^{abc}\,(\chi\chi)\,\partial_a\chi\, \Gamma_b\,\partial_c\chi\right),
\ee
where $\chi\,\chi\equiv \chi^\alpha \,C_{\alpha\beta}\,\chi^\beta\equiv \chi^\alpha\,\chi_\alpha$\,.

The goldstino equation of motion is
\be\label{12eom}
\ii\,\Gamma^a_{\alpha\beta}\,\partial_a\chi^{\beta}=\frac{3f^{-2}}{4}\,\chi_\alpha\,\varepsilon^{abc}\,\partial_a\chi\, \Gamma_b\,\partial_c\chi-\frac{f^{-2}}{2}\,\chi^\gamma\,\Gamma^a_{\gamma\alpha}\,\partial_a\chi\,\Gamma^b\,\partial_b\chi.
\ee
In what follows we will skip the constant term in the action, which however becomes important when the goldstino couples to gravity, since it gives a positive contribution to the cosmological constant.

\subsection{Hamiltonian analysis}
Let us perform the Hamiltonian analysis of this model by determining the form of the Hamiltonian and counting the number of physical degrees of freedom. To this end we split the $D=3$ space-time indices into time and space indices $a=(0,i)$, defining $\varepsilon^{0ij}\equiv \varepsilon^{ij}$ and writing the Lagrangian in the following form:
\be\label{VAa121}
\mathcal L_{1/2}=\ii\,\partial_0\chi\,\Gamma^0\chi-\ii\,\chi\,\Gamma^i\,\partial_i\chi\,
-\frac{f^{-2}}{2}\,\varepsilon^{ij}\,\chi\chi\,\big(\partial_i\chi\,\Gamma_0\,\partial_j\chi-2\,\partial_0\chi\,\Gamma_i\,\partial_j\chi\big)\,.
\ee
The conjugate momentum is
\be\label{p12}
p_\alpha=\frac{\delta L}{\delta \partial_0\chi^\alpha}=\ii\,\Gamma^0_{\alpha\beta}\,\chi^\beta+f^{-2}\,\Gamma_{i\alpha\beta}\,\partial_j\chi^\beta(\chi\chi)
\ee
and the canonical Hamiltonian density is
\be\label{h12}
\mathcal H_{1/2}=\partial_0\chi^\alpha \,p_\alpha-\mathcal L_{1/2}=\ii\,\chi\,\Gamma^i\,\partial_i\chi\,+\frac{f^{-2}}{2}\,\varepsilon^{ij}\,\chi\chi\,\partial_i\chi\Gamma^{0}\,\partial_j\chi\,.
\ee
The canonical anti-commuting Poisson brackets between $\chi^\alpha$ and $p_\beta$ are
\be\label{chip}
\{\chi_{\alpha}(t,\mathbf x),p^\beta(t,\mathbf y)\}=\delta^\beta_\alpha\,\delta({\bf x}-\mathbf y).
\ee

The expression for the momentum \eqref{p12} tells us that it is completely expressed in terms of $\chi$ and its spatial derivatives. Hence, the theory has two constraints,
\be\label{f12}
F_\alpha=p_\alpha-\ii\,\Gamma^0_{\alpha\beta}\,\chi^\beta-f^{-2}\,\varepsilon^{ij}\,\Gamma_{i\alpha\beta}\,\partial_j\chi^\beta(\chi\chi)=0\,,
\ee
which are of the second class in the classification by Dirac \cite{Dirac-lectures,HT-book}, since their equal-time Poisson brackets do not vanish.
\begin{align}\label{pbf12}\nonumber
\{F_{\alpha}(t,\mathbf x),F_\beta(t,\mathbf y)\} =2\,\Big(-\ii\,\Gamma^0_{\alpha\beta}+f^{-2}\,\varepsilon^{ij}\,\big(\,\Gamma_{i\alpha \beta}\,(\chi\,\partial_j\chi)+2\,\partial_j\chi^\rho\,\chi_{(\alpha}\Gamma_{\beta)\rho i}\big)\Big)\,\delta({\bf x}-\mathbf y),
\end{align}
where $\mathbf x$ and $\mathbf y$ stand for the spatial coordinates $x^i$ and $y^i$.

This implies that the goldstino has two independent degrees of freedom in the Hamiltonian phase space (one coordinate and one momentum), and correspondingly a single degree of freedom in the configuration space, i.e. the same as the free Majorana fermion in $D=3$ on the mass-shell.

Let us now evaluate the on-shell value of the quartic-order term in the Hamiltonian \eqref{h12}. To this end, we rewrite this term using gamma-matrix identities, modulo a total derivative, in the following form:
\be\label{4order}
\frac{f^{-2}}{2}\,\varepsilon^{ij}\,\chi\chi\,\partial_i\chi\,\Gamma^{0}\,\partial_j\chi
=\frac{f^{-2}}{2}\,\chi\chi\,\partial_i\chi\,\Gamma^i\,\Gamma^j\,\partial_j\chi-\frac{f^{-2}}{4}\,\partial_i(\chi\chi)\,\partial^i(\chi\chi)
\ee
Now note that the equations of motion \eqref{12eom} imply that
\be\label{12eomc}
\Gamma^i\,\partial_i\chi=-\,\Gamma^0\,\partial_0\chi+\mathcal O\,(\chi\,\partial\chi \,\partial\chi)\,.
\ee
Substituting this expression into \eqref{4order} we get the on-shell value of the Hamiltonian density
\be\label{os12H}
\mathcal H_{1/2}=\ii\,\chi\,\Gamma^i\,\partial_i\chi\,+\frac{f^{-2}}{4}\,\partial_i(\ii\,\chi\chi)\,\partial^i(\ii\,\chi\chi)
+\frac{f^{-2}}{4}\,\partial_0(\ii\,\chi\chi)\,\partial_0(\ii\,\chi\chi)
\ee
in which the quadratic term is the standard free Hamiltonian of a massless Majorana fermion and the quartic terms are manifestly non-negative, since $(\ii\,\chi^\alpha\chi_\alpha)$ is a real (nilpotent) scalar. We have thus verified a well known fact that the higher-order terms in the Volkov-Akulov goldstino model do not bring about unphysical ghost degrees of freedom.

\section{Vector Goldstone model}\label{1}
We now move to the Volkov-Akulov construction of a Goldstone model describing the spontaneous breaking of the rigid symmetry associated with the algebra \eqref{spin1}. In this case the invariant one-form is
\be\label{11}
E^a=dx^a+f^{-2}\,\varepsilon^{abc}A_b(x)\,dA_c(x)
=dx^m(\delta^a_m+f^{-2}\,\varepsilon^{abc}A_b(x)\,\partial_mA_c(x))\equiv dx^mE_m^a\,,
\ee
where $A_a(x)$ is a vector Goldstone  which under  \eqref{spin1} transforms as follows:
\be\label{deltaA}
{x'}^a=x^a-f^{-2}\,\varepsilon^{abc}\,s_b\,A_c(x)\,, \qquad  A'_a(x')=A_a(x)+s_a\,,
\ee
where $s_a$ is a constant vector parameter. The infinitesimal transformation of the form of the goldstone field $A_a(x)$,
\begin{align}\label{vary1}
\delta A_a(x)=s_a+f^{-2}\,\varepsilon^{dbc}\big(s_b\,A_c(x)\big)\,\partial_dA_a(x)\,,
\end{align}
shows that it transforms non-linearly under the symmetry. The commutator of two variations closes on the translation of $A_a$ in accordance with the structure of the algebra  \eqref{spin1},
\begin{align}
    [\delta_2\,,\delta_1]\,A_a(x)=2\,f^{-2}\,\varepsilon^{dbc}\,(s^1_{b}\,s^2_{c})\,\partial_dA_a(x)\,.
\end{align}

\subsection{Action and equations of motion}\label{111}
We construct the action for $A_a(x)$ in the same way as in Section \ref{12}, substituting into \eqref{generic} the one-form \eqref{11}. We thus get (subtracting the constant term and modulo a total derivative),
\bea\label{S1}
S_1&=&-f^{2}\int\,d^3x\,(\det E^a_d-1)\nonumber\\
&=&\int\,d^3x\left(\varepsilon^{abc}A_a\,\partial_bA_c-\frac{f^{-2}}{2}\,\varepsilon^{abc}\varepsilon^{def}\,A_a\,A_d\,\partial_eA_b\,\partial_fA_c\right).
\eea
Note that the quadratic term in \eqref{S1} is the Abelian Chern-Simons action and that the  sixth-order term in $A_a$ (and its derivatives) vanishes.

The equations of motion of $A_a(x)$ which follow from this action have the following form
\begin{align}\label{eom_cs}
\varepsilon^{abc}\,\partial_bA_c
&=f^{-2}\,\varepsilon^{abc}\varepsilon^{def}\,A_d\,\partial_eA_b\,\partial_fA_c\,.
\end{align}
When $f^{-2}=0$, the action and the equations of motion reduce to those of the Chern-Simons theory. In this case the model is invariant under the following gauge transformation of the vector field
\be\label{gauge}
A'_a=A_a+\partial_a\lambda(x),
\ee
and the equations of motion tell us that $A_a(x)$ does not have local physical degrees of freedom. The presence of the gauge symmetry manifests itself in the fact that the Chern-Simons field equations satisfy the Bianchi identity
\be\label{BI}
\partial_a(\varepsilon^{abc}\,\partial_bA_c)\equiv 0.
\ee
When $f^{-2}\not=0$, taking the divergence of the non-linear equation \eqref{eom_cs} we find that, for consistency,
$$
\varepsilon^{abc}\varepsilon^{def}\,\partial_a(A_d\,\partial_eA_b\,\partial_fA_c)=0\,,
$$
but this is not an identity. A possible generalization of the Bianchi identity might be as follows. The equations \eqref{eom_cs} can be expressed in the following form:
\be\label{eom_cs1}
\varepsilon^{abc}\,\mathcal{D}_bA_c=0\,,
\ee
where
\be\label{covdir}
\mathcal{D}_b = (E^{-1})^d_b\,\partial_d + \frac{1}{2\,E}\,\partial_d \left(E\, (E^{-1})^{d}_b\right),
\ee
$(E^{-1})^d_b$ is the matrix inverse of $E^a_b$ defined in \eqref{11} and $E:=\det E^a_b$. One might hope that the operator $\mathcal D_a$  replaces the partial derivative in the sought after generalization of the Bianchi identity \eqref{BI}, but it turns out to not be the case, i.e.
$\varepsilon^{abc}\,\mathcal{D}_a\mathcal{D}_bA_c$ is not identically zero.

If the equations of motion do not satisfy a Bianchi identity (which for generic systems with local symmetries is also known as a Noether identity),
then the non-linear system under consideration is not invariant under a non-linear generalization of the gauge transformation \eqref{gauge} and hence contains propagating degrees of freedom. As a further indication that this is indeed the case let us note that the solution of equations \eqref{eom_cs} can be studied order-by-order in $f^{-2}$ and that it includes a scalar degree of freedom, which is not a pure gauge in the absence of the local symmetry. Indeed, at the zeroth order in $f^{-2}$, the solution of \eqref{eom_cs} is $A^{(0)}_a=\partial_a\varphi$. To order $f^{-2}$ we have
\be\label{1st}
A_a=\partial_a\varphi+f^{-2}A^{(1)}_a+\mathcal O(f^{-4})\,.
\ee
Plugging this into \eqref{eom_cs} we find the expression for the field-strength of $A^{(1)}_a$ in terms of the derivatives of $\varphi$:
\be\label{1st1}
\varepsilon^{abc}\,\partial_bA^{(1)}_c=\varepsilon^{abc}\varepsilon^{def}\,\partial_d\varphi\,\partial_e\partial_b\varphi\,\partial_f\partial_c\varphi\,.
\ee
Upon taking the divergence of the left and right hand sides of \eqref{1st1}), we get,
\bea\label{detphi}
-\frac 16\,\varepsilon^{abc}\varepsilon^{def}\,\partial_a\partial_d\varphi\,\partial_e\partial_b\varphi\,\partial_f\partial_c\varphi &=& \det (\partial_a\partial^b \varphi)\\
&=&(\Box \varphi)^3-3\,\Box\varphi\, \partial_a\partial^b\varphi\,\partial_b\partial^a\varphi+2\,\partial_a\partial^b\varphi\,\partial_b\partial^c\varphi\,\partial_c\partial^a\varphi =0\,.\nonumber
\eea
The latter can be regarded as a higher-order equation of motion of $\varphi$. Note that it is of the second-order in time derivative. This indicates that the model has a scalar propagating degree of freedom. This degree of freedom is of St\"uckelberg type whose equation of motion \eqref{detphi} can be obtained in a proper decoupling limit $f\to\infty$ of a
gauge-invariant action having the same form as \eqref{S1} but in which $A_a$ is replaced with $\hat A_a=A_a-f^{\frac 12}\partial_a\hat\varphi$, the latter being invariant under the field variations $\delta A_a=\partial_a\lambda$ and $\delta\hat\varphi=f^{-\frac 12}\lambda$, and $f^{\frac 12}\hat\varphi=\varphi$.
In the decoupling limit the Lagrangian reduces to
\be\label{deca}
{\mathcal L}(\hat A_a)|_{f\to\infty}=\varepsilon^{abc}\,A_a\,\partial_bA_c-\frac 12 \,\varepsilon^{abc}\varepsilon^{def}\,\partial_a\hat\varphi\,\partial_d\hat\varphi\,\partial_e\partial_b\hat\varphi\,\partial_f\partial_c\hat\varphi\,.
\ee
The  field $\hat\varphi$ is of mass dimension $M^{-\frac 34}$, which is not canonical. We can introduce the scalar field with the canonical mass dimension  $M^{\frac 12}$ by rescaling $\hat\varphi \to M^{-\frac 54}\hat\varphi$. This results in the appearance of the coupling constant $M^{-5}$ in the Lagrangian.

Upon integrating by parts, we can bring the scalar part of this Lagrangian to the following form
\bea\label{deca1}
{\mathcal L}(\hat \varphi)&=&\frac {M^{-5}}2\, \hat\varphi\,\varepsilon^{abc}\varepsilon^{def}\,\partial_a\partial_d\hat\varphi\,\partial_e\partial_b\hat\varphi\,\partial_f\partial_c\hat\varphi\nonumber\\
&=&-\,3\,M^{-5} \hat \varphi\,\det(\partial_a\partial^b\hat\varphi)\nonumber\\
&=&-\,\frac{M^{-5}}2\hat \varphi\,\big((\Box \hat\varphi)^3-3\,\Box\hat\varphi\, \partial_a\partial^b\hat\varphi\,\partial_b\partial^a\hat\varphi+2\,\partial_a\partial^b\hat\varphi\,\partial_b\partial^c\hat\varphi\,\partial_c\partial^a\hat\varphi)\big)\,.
\eea
Curiously, the form \eqref{deca1} of the higher-order scalar Lagrangian is the same as the quartic term in the Galileon Lagrangian \cite{Nicolis:2008in} and a corresponding term in a (beyond) Horndeski tensor-scalar theory of gravity \cite{Horndeski:1974wa,Gleyzes:2014dya}. The Lagrangian is invariant (modulo a total derivative) under the Galileon symmetry transformations $\hat\varphi \to \hat\varphi+c+c_ax^a$, where $c$ and $c_a$ are constant parameters.

The equation of motion of $\hat\varphi$ which follows from this Lagrangian is eq. \eqref{detphi}. Simplest non-trivial solutions of this equation are the static fields
\be\label{static}
\partial_t \hat\varphi(t,x^i)=0,
\ee
and plain-wave-like solutions
\be\label{plain}
\hat \varphi=e^{ip_ax^a}\phi(p)+e^{-ip_ax^a}\phi^*(p),
\ee
where $p_a$ is an arbitrary time-like, space-like or light-like momentum. It is a priori not subject to the mass-shell condition $p_ap^a-m^2=0$ since the Lagrangian does not contain the quadratic kinetic term $\mathcal L_2=-\,\frac 12\,(\partial_a\hat\varphi\partial^a\hat\varphi+m^2\hat\varphi^2)$. Hence, there is no corresponding term in the equation of motion. So, this higher-order model contains tachyons, unless they are excluded by imposing appropriate mass-shell conditions on $\hat\varphi$.

To prove that $\hat\varphi$ is the only propagating mode in this model and to further study its dynamical properties we now move to the Hamiltonian analysis.

\subsection{Hamiltonian analysis of the vector Goldstone model}
Splitting the space-time indices, we rewrite the action \eqref{S1} in the following form
\bea\label{S1h}
\mathcal S_1 &=&\int\,d^3x\,\varepsilon^{ij}\,(2A_0 \,\partial_i A_j+A_j \,\partial_0 A_i)  \\
    &&+f^{-2}\int \,d^3x\,\varepsilon^{ij}\varepsilon^{kl}\Big(A_j A_k(\partial_iA_l\,\partial_0A_0 -\partial_lA_0 \,\partial_0 A_i)\nonumber\\
    &&{}\qquad\qquad\qquad +A_0 A_k\, \partial_0A_i \,\partial_l A_j -A_0^2 \,\partial_lA_j \,\partial_kA_i \Big).\nonumber
\eea
Note that, as in the fermionic case, the action is of the first order in time derivative. Hence, the canonical momenta $p^a=\frac{\delta\mathcal L}{\delta (\partial_0A_a)}$ are expressed in terms of the components of $A_a$ and their spatial derivatives. Thus we get three primary constraints:
\bea\label{Ci}
C^i=p^i-\varepsilon^{ij}A_j+f^{-2}\varepsilon^{ij}\varepsilon^{kl}(A_j\,A_{k}\,\partial_{l}A_0-A_0\,A_k\,\partial_{l}A_{j})=0\,,
\eea
\be\label{C0}
C^0=p^0-f^{-2}\,\varepsilon^{ij}\varepsilon^{kl}\,A_j\, A_k\,\partial_iA_l=p^0-f^{-2}\,\varepsilon^{ij}\varepsilon^{kl}\,A_j\, A_k\,\partial_lA_i=0\,.
\ee
The canonical (equal-time) Poisson brackets of $A_a$ and $p^b$ are
\be\label{Ap}
[A_a(t,\mathbf x),p^b(t,\mathbf y)]=\delta^b_a\,\delta(\mathbf x-\mathbf y),\qquad (a=0,i)\,.
\ee
We find that the Poisson brackets of $C^i$ do not vanish on the constraint surface
\be\label{C1C2}
[C^i(t,\mathbf x),C^j(t,\mathbf y)]=-\,2\,\varepsilon^{ij}\Big(1-2\,f^{-2}\,\varepsilon^{kl}A_{k}\,\partial_{l}A_0(\mathbf x)\Big)\delta(\mathbf x-\mathbf y)\,,
\ee
and
\be\label{CiC0}
[C^i(t,\mathbf x),C^0(t,\mathbf y)]=-4\,f^{-2}\,\varepsilon^{ij}\varepsilon^{kl}A_k\,\partial_lA_j\,\delta(\mathbf x-\mathbf y)\,.
\ee
Constraint \eqref{C0} can be modified as following to make it commute with $C^i\,$:
\be\label{hatC0}
\hat C^0=C^0-\frac{2\,f^{-2}\varepsilon^{kl}A_k\,\partial_l A_i}{1-2\,f^{-2}\,\varepsilon^{kl}A_{k}\,\partial_{l}A_0}\,C^i\,.
\ee
Therefore, the constraints $C^i$ are of the second class according to the classification by Dirac. They are not associated with gauge symmetries of the model. We now turn to the identification of secondary constraints. To this end, following Dirac formalism, we construct the Hamiltonian which includes the canonical Hamiltonian and the primary constraints multiplied by Lagrange multipliers,
\be\label{HT}
H_T=\int d^2{\mathbf y}\,(\mathcal H_c+u_i\,C^i+u_0\,\hat C^0)\,,
\ee
where
\be\label{Hd}
\mathcal H_c=p^aA_a-\mathcal L_1=-\,2\,A_0\,\varepsilon^{ij}\,\partial_iA_j+f^{-2}\,\varepsilon^{ij}\varepsilon^{kl}\,(A_0)^2\,\partial_iA_k\,\partial_{j}A_{l}\,.
\ee
The consistency (i.e. time-independence) of the constraints requires that the Poisson brackets of the constraints with the Hamiltonian \eqref{HT} vanish. For the constraints $C^i$ this requirement fixes the value of the Lagrange multipliers $u_i(x)$, while the requirement of the vanishing of the Poisson bracket of $H_T$ with $\hat C^0$ produces the secondary constraint
\begin{align}\label{B}
&[\hat C^0,H_T]=0 \nonumber \\
& \Rightarrow B=\varepsilon^{ij}\,\partial_iA_j-f^{-2}A_0\,\varepsilon^{ij}\varepsilon^{kl}\,\partial_{k}A_{i}\,\partial_lA_j\,
-2\,f^{-2}\varepsilon^{ij}\varepsilon^{kl}A_j\,\partial_lA_i\,\partial_kA_0\,=0\,.
\end{align}
The Poisson bracket of $B$ with $C^i$ is
\bea\label{CiB}
[B(\mathbf x),C^i(\mathbf y)]
\if{}
=-\,\varepsilon^{ij}\,\partial_{x^j}\delta({\bf x}-{\bf y})+2\,f^{-2}\,\varepsilon^{ij}\varepsilon^{kl}\,
\partial_kA_0(t,{\bf x})\,\partial_l A_j(t,{\bf x})\delta({\bf x}-{\bf y})+2\,f^{-2}\,\varepsilon^{ij}\varepsilon^{kl}\,\bigg(\,A_0(t,{\bf x})\,\partial_kA_j(t,{\bf x})-A_j(t,{\bf x})\,\partial_k A_0(t,{\bf x}) \bigg)\partial_{x^l}\delta({\bf x}-{\bf y})\nonumber\\
\fi
&=&-\,\varepsilon^{ij}\,\partial_{x^j}\delta({\bf x}-{\bf y})+6\,f^{-2}\,\varepsilon^{ij}\varepsilon^{kl}\,
\partial_kA_0(t,{\bf x})\,\partial_l A_j(t,{\bf x})\,\delta({\bf x}-{\bf y})\\
&&+2\,f^{-2}\,\varepsilon^{ij}\varepsilon^{kl}\,\partial_{x^l}\bigg((\,A_0(t,{\bf x})\,\partial_kA_j(t,{\bf x})-A_j(t,{\bf x})\,\partial_k A_0(t,{\bf x}))\,\delta({\bf x}-{\bf y})\bigg)\,.\nonumber
\eea
We can make this Poisson bracket vanish by modifying the constraint $B$ \eqref{B} as follows
\be\label{hatB}
\hat B=B-6\,f^{-2}\,\varepsilon^{kl}\,
\partial_kA_0\,\partial_l A_j\,\hat C^j+\partial_j{\hat C^j}-2f^{-2}\varepsilon^{kl}\partial_l\bigg((A_0 \,\partial_kA_j-A_j\partial_k A_0)\,\hat C^j\bigg),
\ee
where
\be\label{hatCj}
\hat C^j=\frac{C^j}{2\,(1-2\,f^{-2}\varepsilon^{kl}A_{k}\,\partial_{l}A_0)}\,,\qquad [\hat C^j(t,\mathbf x), C^i(t,\mathbf y)]=\varepsilon^{ij}\,\delta({\bf x}-{\bf y})\,.
\ee
Thus
\be\label{hatBCi}
[\hat B,C^i]=0\,.
\ee
However, $B$ has (in general) a non-vanishing Poisson bracket with $C^0$
\be\label{hatBhatC0}
[B(\mathbf x),C^0(\mathbf y)]=-f^{-2}\varepsilon^{ij}\varepsilon^{kl}\,\partial_{k}A_{i}\,\partial_lA_j\,\delta({\bf x}-{\bf y})\,
-2f^{-2}\varepsilon^{ij}\varepsilon^{kl}A_j\,\partial_lA_i\,\partial_{x^k}\delta({\bf x}-{\bf y})\,.
\ee
If we take the linear combination of the constraints $ B$ and $C^0$, namely $B_1=\frac 12\,(B+C^0)$ and $B_2=\frac 12\,(B-C^0)$, the Poisson bracket simplifies to
\be\label{hatBhatC01}
[B_1(\mathbf x),B_2(\mathbf y)]=f^{-2}\varepsilon^{ij}\varepsilon^{kl}\,\partial_{k}A_{i}\,\partial_lA_j\,\delta({\bf x}-{\bf y}).
\ee
The Poisson bracket \eqref{hatBhatC01} vanishes when $f^{-2}=0$, i.e. in the case of free Chern-Simons theory. Then the constraints $C^0$ and $B$ (or equivalently $B_1$ and $B_2$) are of the first class. They generate the local symmetry of the Chern-Simons action, which implies that the CS vector field does not have propagating degrees of freedom. Indeed, in the Hamiltonian formulation $A_a$ and its conjugate momenta $p^a$ have 3+3=6 components. These are related to each other by the second-class constraints \eqref{Ci} which remove two degrees of freedom. The two first class constraints, remove two degrees of freedom each, i.e. 4, and hence there is no physical degree of freedom left. Note that in this case the Hamiltonian \eqref{HT} is zero on the constraint surface, which also points at the absence of propagating modes.

In the non-linear case in which $f^{-2}\not =0$, the Poisson bracket \eqref{hatBhatC01} is non-zero for a generic field  $A_a$, therefore the constraints $ C^0$ and $B$ become of the second-class and remove only two degrees of freedom.
 One can also check that the non-linear model does not have tertiary constraints, i.e. that the Poisson brackets of the primary and the secondary constraints with the Hamiltonian \eqref{HT} vanish provided the Lagrange multipliers $u_i$ and $u_0$ are appropriate functions of $A_a$ and its derivatives.
 We are thus left with two Hamiltonian degrees of freedom contained in $A_a$ and $p^a$, which correspond to a single degree of freedom in the Lagrangian formulation. This is the scalar mode discussed at the end of Section \ref{111}.

 To elucidate the physical properties of this mode, let us look at the form of the Hamiltonian, eqs. \eqref{HT} and \eqref{Hd}, in the non-linear case. We see that the Hamiltonian density \eqref{Hd} does not vanish on the constraint surface anymore. Modulo the constraint \eqref{B} and up to a total derivative, it has the following form:
 \be\label{Hd1}
 \mathcal H_c=-\,3\,f^{-2}\,(A_0)^2\,\varepsilon^{ij}\varepsilon^{kl}\,\partial_iA_k\,\partial_jA_{l}\equiv -\,6\,f^{-2}\,(A_0)^2\det \partial_iA_j\,.
 \ee
Note that this Hamiltonian density is non-zero for the perturbative solution \eqref{1st}-\eqref{detphi}, and it is not bounded from below for generic classical values of the field $A_a$, since $\det \partial_iA_{j}$ is not positive definite. In the decoupling limit \eqref{deca} it reduces to the Hamiltonian density for the St\"uckelberg field $\hat\varphi(x)$
\be\label{Hphi}
 \mathcal H_{\hat\varphi}= -\,\frac{p_{\hat\varphi}^2}{6\det \partial_i\partial_j\hat\varphi}\,, \qquad p_{\hat\varphi}=-\,6\,(\det \partial_i\partial_j\hat\varphi) \,\partial_0\hat\varphi\,.
 \ee
Equation \eqref{Hphi} is the three-dimensional counterpart of the quartic Galileon term in the Hamiltonian of the generic $D=4$ Galileon theory derived in \cite{Zhou:2010di,Sivanesan:2011kw}.

Let us look at the value of this Hamiltonian for fluctuations around a simple static solution  $\hat\varphi_0=\frac 12\, x^ix^i$, $p_{\hat\varphi}=0$ (whose Hamiltonian, and hence energy, is zero)
\be
\hat\varphi=\hat\varphi_0+\delta\phi.
\ee
Then, to the second order in $\delta\phi$ we have
\be
\mathcal H_{\delta\phi}=-\,\frac {p^2_{\delta\phi}}{6}=-\,6\,\delta \dot\phi^2,
\ee
which is negative.

Note that if we changed the sign of the initial Lagrangian in \eqref{S1} (which a priori is equally admissible, since the Chern-Simons term may have any sign), we would get the Hamiltonian with the plus sign in \eqref{Hd1} and \eqref{Hphi}. Then the quadratic Hamiltonian density of the fluctuations around the classical solution above would be positive. But if instead, we consider fluctuations around zero-energy static solutions, e.g. of the form  $\hat\varphi_0=e^{a_ix^i}b+{\rm c.c.}$ (where $a_i$ and $b$ are  complex constants), their Hamiltonian density would be negative.

To summarize, the vector Goldstone model describing the spontaneous breaking of the rigid symmetry generated by the algebra \eqref{spin1} does not maintain the local gauge symmetry of the quadratic Chern-Simons action. Due to the presence of the non-linear terms in the action there is a propagating scalar degree of freedom whose Hamiltonian is not bounded from below. This, in general, makes this model classically unstable, even though the Lagrangian is linear in the time derivative of $A_a(x)$.

\section{Spin-3/2 goldstino model}\label{s32}

In the spin-3/2 case the action for a Goldstone field $\chi^\alpha_a(x)$ associated with the spontaneous breaking of spin-$\frac 32$ supersymmetry generated by \eqref{spin321} is constructed with the use of the one-form
\be\label{Eam}
E^a=dx^dE^{\,a}_d=dx^d\,(\delta^a_d+\ii f^{-2}\,\varepsilon^{abc}\,\chi_b\,\partial_d\chi_c)\,,
\ee
which is invariant under the following variations of $x^a$ and $\chi^\alpha_a(x)$
\be\label{vary321}
{x'}^a=x^a-\ii \,f^{-2}\,\varepsilon^{abc}\,\zeta^\alpha_b\,\chi_{\alpha c}\,, \qquad {\chi'}^\alpha_a(x')=\chi^\alpha_a(x)+\zeta^\alpha_a\,,\nonumber
\ee
\be\label{vary32}
\delta\chi^\alpha_a(x)=\zeta^\alpha_a+\ii\,f^{-2}\varepsilon^{dbc}\,\big(\zeta_b\,\chi_{c}(x)\big)\,\partial_d\chi^\alpha_a(x)\,,
\ee
where $\zeta^\alpha_a$ is a constant parameter. Note that, as for all the other cases, the commutator of two variations \eqref{vary32} closes on the translations off the mass shell, i.e. without the use of the equations of motion:
\be
[\delta_2,\delta_1]\,\chi_a^\alpha=\xi^d\,\partial_d\chi_a^\alpha\,,  \qquad \xi^d=2\,\ii\,f^{-2}\,\varepsilon^{dbc}\,\zeta^1_b\,\zeta^2_{c}\,.
\ee

The spin-3/2 goldstino action has the following form
\bea\label{VA32}
S_{3/2}&=&-f^{2}\int d^3x \,(\det E^a_d-1)\nonumber\\
&=&\int d^3x \left(\ii\,\varepsilon^{abc}\,\chi_a\,\partial_b\chi_c+
\frac{f^{-2}}{2}\,\varepsilon^{abc}\varepsilon^{dfg}\,\big((\chi_a\,\partial_b\chi_c)\,(\chi_d\,\partial_f\chi_g)
-(\chi_b\,\partial_d\chi_c)\,(\chi_f\,\partial_a\chi_g)\big)\right.
\nonumber\\
& &\left.+\frac{\ii\,f^{-4}}{6}\,\varepsilon^{a'b'c'}\,
(\varepsilon^{abc}\varepsilon^{def}-\varepsilon^{abf}\varepsilon^{dec})
 \,(\chi_c\,\partial_{a'}
\chi_f)\,(\chi_a\,\partial_{b'}\chi_b)\,(\chi_d\,\partial_{c'}\chi_e)\right).
\eea
and the equations of motion have the form similar to \eqref{eom_cs1}
\be\label{32eom}
\varepsilon^{abc}\,{\mathcal D}_b\chi_c=0\,.
\ee
We see that the quadratic term in the action \eqref{VA32} is the action for a $D=3$ (Rarita-Schwinger) spin-$\frac 32$ free massless field which is invariant under conventional (linearized) local supersymmetry variations $\delta\chi^\alpha_a=\partial_a\epsilon^\alpha(x)$.
Let us figure out if in contrast to the spin-1 case, the spin-3/2 goldstino action can be invariant under a non-linear generalization of this symmetry.
Again, let us first look at what happens with the model if we use the St\"uckelberg trick and take a limit $f\to\infty$. To this end we replace in the action \eqref{VA32} the field $\chi_a$ with its gauge-invariant counterpart $\hat \chi_a=\chi_a+f^{\frac 23}\partial_a \psi$, where $\psi$ is the St\"uckelberg spinor field and the normalization with the factor $f^{2/3}$ is chosen to perform a certain limit $f\to\infty$ in the action. By construction $\hat\chi_a$ is invariant under the gauge transformations $\delta\chi_a=\partial_a\epsilon(x)$, $\delta\psi=-f^{-\frac 23}\epsilon(x)$ which can be used to completely eliminate the latter. On the other hand, sending $f\to\infty$ we obtain the following limit of the model in which however $\chi_a$ and $\psi$ do not decouple from each other
\bea\label{VA32deca}
S_{f\to \infty}
&=&\int d^3x \left(\ii\,\varepsilon^{abc}\,\chi_a\,\partial_b\chi_c+2\,\varepsilon^{abc}\varepsilon^{dfg}
(\chi_a\,\partial_d\partial_c\psi)\,(\partial_f\psi\,\partial_b\partial_g\psi)
-\frac 13\tr(M^3)\right),
\eea
where $M^{a}{}_d=\ii\,\varepsilon^{abc}\,\partial_b\psi\,\partial_d\partial_c\psi$.

Note that in contrast to the vector-field case in which in the decoupling limit the Lagrangian for the St\"uckelberg scalar field is that of the quartic Galileon, see eq. \eqref{deca1}, in the present case the quartic term
$$
\varepsilon^{abc}\varepsilon^{dfg}\,
(\partial_b\psi\,\partial_d\partial_c\psi)\,(\partial_f\psi\,\partial_a\partial_g\psi)
=\partial_b\left(\varepsilon^{abc}\varepsilon^{dfg}(\psi\,\partial_d\partial_c\psi)\,(\partial_f\psi\,\partial_a\partial_g\psi)\right)
$$
is a total derivative, since
\be\label{0}
\varepsilon^{abc}\varepsilon^{dfg}(\partial_c\partial_d\psi^\alpha)\,(\partial_b\partial_f\psi\,\partial_a\partial_g\psi)\equiv 0\ee
due to the anti-commutativity of $\psi$ and the total symmetry of this expression in the exchange of the pairs of the indices $cd,bf$ and $ag$. This term can thus be discarded, and there is no decoupling limit of the spin-3/2 action similar to that of the vector-field model. The triviality of this term also implies that the quartic term in the action \eqref {VA32} vanishes (modulo a total derivative) on the
solution of the free Rarita-Schwinger field equation. Notice also that the action \eqref{VA32deca} is invariant under the gauge transformation $\delta\chi_a=\partial_a\lambda(x)$, $\delta\psi=0$ due to the same identity \eqref{0}.

The equation of motion of $\chi_a$, which follows from \eqref{VA32deca}, is
\be\label{eochia}
\varepsilon^{abc}\,\partial_b\chi^\alpha_c=\ii\,\varepsilon^{abc}\,\varepsilon^{dfg}\,
\partial_d\partial_c\psi^\alpha\,(\partial_f\psi\,\partial_b\partial_g\psi)\,.
\ee
Using the identity
\bea\label{id}
\varepsilon^{abc}\,\varepsilon^{dfg}\,
\partial_d\partial_c\psi^\alpha\,(\partial_f\psi\,\partial_b\partial_g\psi)&\equiv & \frac 12\,\varepsilon^{abc}\varepsilon^{dfg}\,
(\partial_f\partial_c\psi\,\partial_b\partial_g\psi)\,\partial_d\psi^\alpha\nonumber\\
&&
\equiv- \,\frac 13\,\varepsilon^{abc}\,\varepsilon^{dfg}\,
\partial_b\left(\partial_d\psi^\alpha\,(\partial_f\psi\,\partial_c\partial_g\psi)\right)
\eea
we find that the general solution of \eqref{eochia} is
\be\label{gs}
\chi^\alpha_c=\partial_c\epsilon^\alpha-\frac{\ii}3\,\varepsilon^{dfg}\,\partial_d\psi^\alpha\,(\partial_f\psi\,\partial_c\partial_g\psi)\,.
\ee
This implies that, modulo the pure gauge degree of freedom, the field $\chi_a$ is completely determined in terms of derivatives of $\psi$. As can be verified, the equations of motion of $\psi$ which follow from \eqref{VA32deca} are identically satisfied, and hence $\psi$ is completely arbitrary in this limit. Moreover, action \eqref{VA32deca} can be recast into the Chern-Simons form as following:
\be\label{hatCS}
S_{f\to \infty}
=\ii\,\int d^3x \,\varepsilon^{abc}\Big(\chi^\alpha_a+\frac{\ii}3\varepsilon^{dfg}\,\partial_d\psi^\alpha\,(\partial_f\psi\,\partial_a\partial_g\psi)\Big)\partial_b\Big(\chi_{c\alpha}+\frac{\ii}3\,\varepsilon^{pqr}\,\partial_p\psi_\alpha\,(\partial_q\psi\,\partial_c\partial_r\psi)\Big).
\ee
This action  turns out to be invariant under the following gauge symmetry transformation
\bea\label{nls}
\delta \psi&=&\epsilon(x),\nonumber\\
\delta\chi^\alpha_a &=&\partial_a\lambda^\alpha(x)
-\frac{\ii}3\,\varepsilon^{dfg}\,\Big(\partial_d\epsilon^\alpha\,(\partial_f\psi\,\partial_a\partial_g\psi)
+\partial_d\psi^\alpha\,(\partial_f\epsilon\,\partial_a\partial_g\psi)
+\partial_d\psi^\alpha\,(\partial_f\psi\,\partial_a\partial_g\epsilon)\Big)\nonumber\\
&\equiv&\partial_a\Big(\lambda^\alpha(x)-\frac{\ii}3\varepsilon^{dfg}\,\partial_d\psi^\alpha\,(\partial_f\psi\,\partial_g\epsilon)\Big)
-\ii\,\varepsilon^{dfg}\,(\partial_d\epsilon\,\partial_a\partial_f\psi)\,\partial_g\psi^\alpha\,,
\eea
where $\lambda^\alpha(x)$ and $\epsilon^\alpha(x)$ are independent parameters. Hence, $\psi$ is a pure gauge.

Note also that the above analysis actually prompts us the form of the perturbative solution of the full non-linear equation of motion \eqref{32eom} up to the order $f^{-2}$. It is obtained from \eqref{gs} by re-scaling $\psi\to f^{-\frac 23}\psi$ and taking $\epsilon=\psi$:
\be\label{f-2}
\chi^\alpha_a=\partial_a\psi^\alpha-\frac{\ii f^{-2}}3\,\varepsilon^{dfg}\partial_d\psi^\alpha\,(\partial_f\psi\partial_a\partial_g\psi)
+\mathcal O(f^{-4})\,.
\ee

Moreover, the non-linear symmetry in this limit and the form of the action \eqref{hatCS} prompt us that the full action \eqref{VA32} can be written as following:
\be\label{hatCS1}
S_{3/2}
=\ii\,\int d^3x \,\varepsilon^{abc}\Big(\chi^\alpha_a+\frac{\ii f^{-2}}3\varepsilon^{dfg}\chi_d^\alpha\,
(\chi_f\partial_a\chi_g)\Big)\partial_b\Big(\chi_{c\alpha}+\frac{\ii f^{-2}}3\varepsilon^{pqr}\chi_{p\alpha}\,(\chi_q\partial_c\chi_r)\Big).
\ee
Indeed, \eqref{hatCS1} and \eqref{VA32} are equal to each other modulo a total derivative due to the following identities:
$$
\varepsilon^{abc}\,\varepsilon^{dfg}(\chi_c\,\chi_d)(\partial_b\chi_f\,\partial_a\chi_g)=-\,2\,\varepsilon^{abc}\,\varepsilon^{dfg}\,(\chi_b\,\partial_c\chi_d)(\chi_f\,\partial_a\chi_g),
$$
$$
\varepsilon^{abc}\,\varepsilon^{dfg}\,\varepsilon^{pqr}\,(\chi_f\,\partial_a\chi_g)(\chi_d\,\chi_p)(\partial_b\chi_q\,\partial_c\chi_r)
=2\,\varepsilon^{abc}\,\varepsilon^{dfg}\,\varepsilon^{pqr}\,(\chi_f\,\partial_a\chi_g)(\chi_d\,\partial_b\chi_p)(\chi_q\,\partial_c\chi_r)\,.
$$
The action \eqref {hatCS1} reduces to the free Rarita-Schwinger action
\be\label{RS}
S_{RS}=\ii\,\int d^3x \,\varepsilon^{abc}\,\hat\chi_a\,\partial_b\hat\chi_c
\ee
upon the following field redefinition
\be\label{frd}
\hat\chi_a^\alpha=\chi^\alpha_a+\frac{\ii f^{-2}}3\,\varepsilon^{dfg}\,\chi_d^\alpha\,
(\chi_f\,\partial_a\chi_g).
\ee
This equation is invertible, and using an iteration procedure one can find an explicit expression for $\chi_a$ as a polynomial in $\hat\chi_a$ and $\partial_b\hat\chi_a$,  which stops at most at the sixth order in $\hat\chi$, because of the nilpotency of the latter. Thus up to the order $f^{-4}$, we get,
\bea\label{chi=hat}
\chi_a^\alpha &=&\hat\chi^\alpha_a-\frac{\ii f^{-2}}3\,\varepsilon^{dfg}\,\hat\chi_d^\alpha\,
(\hat\chi_f\,\partial_a\hat\chi_g)\\
&& -\,\frac{ f^{-4}}3\,\varepsilon^{dfg}\,\varepsilon^{pqr}\Big(\hat\chi_g^\alpha\,(\hat\chi_q\,\partial_d\hat\chi_r)
(\hat\chi_p\,\partial_a\hat\chi_f)+\frac{ 1}3\,\partial_a\Big(\hat\chi^\alpha_d(\hat\chi_f\,\hat\chi_p)(\hat\chi_q\,\partial_g\hat\chi_r))\Big)+\mathcal O(f^{-6})\,.\nonumber\
\eea

Action \eqref{hatCS1} (and \eqref{VA32}) is invariant under the following gauge transformation:
\be\label{dghat}
\delta \hat\chi_a^\alpha=\partial_a\epsilon^\alpha=\delta\chi^\alpha_a+\frac{\ii f^{-2}}3\,\varepsilon^{dfg}\,\partial_a\Big(\chi_d^\alpha\,(\chi_f\,\delta\chi_g)\Big)+{\ii\, f^{-2}\,} \varepsilon^{dfg}\,(\delta\chi_d\,\partial_a\chi_f)\,\chi^\alpha_g\,,
\ee
from which by the same iteration procedure one can get the gauge variation of $\chi_a$:
\be\label{varchi}
\delta\chi_a^\alpha=\partial_a\Big(\epsilon^\alpha-\frac{\ii f^{-2}}3\,\varepsilon^{dfg}\,\chi_d^\alpha\,(\chi_f\,\partial_g\epsilon)\Big)-{\ii f^{-2}\,} \varepsilon^{dfg}\,(\partial_d\epsilon\,\partial_a\chi_f)\,\chi^\alpha_g+\mathcal O(f^{-4})\,.
\ee
It is instructive to notice that the commutator of two transformations \eqref{varchi} is exactly zero (to all orders)
\be\label{e1e2}
[\delta_{\epsilon_1},\delta_{\epsilon_2}]\,\chi_a^\alpha\equiv 0\,.
\ee
By construction, the action \eqref{VA32} and hence \eqref{RS} are also invariant under the rigid spin-3/2  supersymmetry variations of the goldstino $\chi_a$ \eqref{vary32} with the corresponding variations of $\hat\chi_a$  derived from \eqref{frd} being of the following form
\bea\label{hatchic}
&\delta\hat\chi^\alpha_a=\zeta^\alpha_a+\ii\,f^{-2}\,\varepsilon^{dbc}\,(\zeta_b\,\hat\chi_{c})\,\partial_d\hat\chi^\alpha_a+\frac{\ii f^{-2}}3\,\varepsilon^{dbc}\,
\Big((\hat\chi_b\,\partial_a\hat\chi_c)\,\zeta_d^\alpha+(\zeta_b\,\partial_a\hat\chi_c)\hat\chi_d^\alpha\Big)+\mathcal O(f^{-4})\,,&\nonumber\\
&[\delta_2,\delta_1]\,\hat\chi_a^\alpha\equiv\xi^d\,\partial_d\hat\chi_a^\alpha\,,  \qquad \xi^d=2\,\ii\,f^{-2}\,\varepsilon^{dbc}\,\zeta^1_b\,\zeta^2_{c}\,.&
\eea
We have thus found that the free Rarita-Schwinger action \eqref{RS} is non-manifestly invariant under the rigid spin-3/2 supersymmetry with the Rarita-Schwinger field being its goldstino transforming non-linearly under the symmetry as in \eqref{hatchic}.

\section{Conclusion and outlook}
We have found that the simplest examples of spontaneous breaking of symmetries introduced by Hietarinta \cite{Hietarinta:1975fu} and the corresponding Goldstone models are specific non-linear generalizations of the Chern-Simons and Rarita-Schwinger Lagrangians.

In the vector algebra case, the spontaneous breaking of the rigid symmetry leads to the breaking of the gauge symmetry of the Abelian Chern-Simons action. As a result, the Chern-Simons Goldstone propagates a scalar mode which turns out to be a Galileon field that appears in the theories of modified gravity. In this respect it would be of interest to consider the coupling of the Chern-Simons Goldstone to a $3d$ gravity model which is invariant under the local symmetry associated with the algebra \eqref{spin1}. As we mentioned in the Introduction, the algebra \eqref{spin1} is a contraction of $so(2,2) = sl(2,\mathbf R) \oplus sl(2,\mathbf R)$ on which the Chern-Simons description of the conventional $3d$ gravity is based \cite{Achucarro:1987vz,Witten:1988hc}.
But the full algebra also includes the Lorentz generators \eqref{vector}.
Therefore, our $3d$ gravity model will contain two spin-2 gauge fields, the conventional gravity dreibein $e^a(x)=dx^me_m^a(x)$ associated with $P_a$ and a dreibein $f^a(x)=dx^mf_m^a(x)$ associated with $S_a$, as well as the spin connection $\omega^a(x)=dx^m\omega^a_m(x)$ associated with the Lorentz generators $M_a=\frac 12\epsilon_{abc}M^{bc}$. An action for these (a priori) independent fields, which is invariant under the local symmetries \eqref{vector} and \eqref{spin1}, has the following form
\be\label{zd}
S=\int (e^a\wedge R_a+\frac 12f^a\wedge Df_a),
\ee
where $R^a=d\omega^a+\frac12 \epsilon^{abc}\omega_b\wedge \omega_c$ is the curvature and $Df_a=df_a+\varepsilon_{abc}\omega^b\wedge f^c$ is the covariant derivative associated with the local Lorentz transformations. The local symmetry variations of the fields are
$$
\delta e^a=D\xi^a(x)+\epsilon^{abc}e_b\lambda_c(x)+\varepsilon^{abc}f_bs_c(x),
$$
\be\label{tr}
\delta f^a=Ds^a(x)+\epsilon^{abc}f_b\lambda_c(x), \qquad \delta \omega^a =D\lambda^{a}(x),
\ee
where $\xi^a(x)$, $s^a(x)$ and $\lambda^a(x)$ are the parameters associated with the generators $P_a$, $S_a$ and $M_a$, respectively. It is easy to see, by analysing the equations of motion, that all the gauge fields in this model are non-dynamical \footnote{Note that the action \eqref{zd} is straightforwardly generalized to describe the similar coupling to gravity of higher-spin fields. To this end one should just promote the one-form field $f^a(x)$ and the gauge parameter $s^a(x)$ to (generically mixed-symmetry) tensors $f^{ab_1\ldots b_n}$ and $s^{ab_1\ldots b_n}$, and appropriately adjust  the contraction of the indices and the Lorentz transformations of $f^{ab_1\ldots b_n}$ in \eqref{zd}-\eqref{tr}.}.

What kind of $3d$ massive gravity or bi-gravity will one obtain when the Goldstone $A_a(x)$ is coupled to \eqref{zd} and generates a Higgs effect? Will it have a relation to one of the three-dimensional gravity models considered in \cite{Bergshoeff:2009hq,Bergshoeff:2009aq,Bergshoeff:2013xma,Ozkan:2018cxj}? We will address these questions in a separate work.

In contrast to the vector Chern-Simons case, in the spin-3/2 goldstino model, upon a non-linear field redefinition,  the free Rarita-Schwinger action itself turns out to be non-manifestly invariant under
the rigid spin-3/2 spersymmetry \eqref{spin321}  which is non-linearly realized on the variations of the Rarita-Schwinger goldstino \eqref{hatchic}. In the presence of the couplings of the spin-3/2 goldstino to other fields, the non-linear field redefinition may no longer remove the non-linear terms and the two forms of the spin-3/2 goldstino models may not be equivalent anymore. In this respect,
it would be of interest to couple the Rarita-Schwinger goldstino to other matter and gauge fields such as (super)gravity and Hypergravity with spin-2 and spin-5/2 gauge fields and to study the properties of these models.

Another interesting problem is to consider a four-dimensional Rarita-Schwinger goldstino model associated with the following algebra:
\begin{equation}\label{4d32}
\{Q_\alpha^a,Q^b_\beta\}=2\,\varepsilon^{abcd}\,(\Gamma_5\,\Gamma_c)_{\alpha\beta}\,P_d \quad (\alpha,\beta=1,...,4),
\quad (a,b,...=0,1,2,3),
\end{equation}
to figure out whether also in this case the non-linear Lagrangian is related to the free Rarita-Schwinger Lagrangian upon a non-linear field redefinition and see whether the non-linearly realized symmetry \eqref{4d32} may fit into the formulation of $N=1, D=4$ supergravity as a non-linear realization of two complex finite-dimensional supergroups considered in \cite{Ivanov:1984nu,Ivanov:1992ax,Maxera:1994xv}.

One can also look at the generalizations of the construction considered in this paper for studying higher-spin Goldstone models.

\subsection*{Acknowledgements}
The authors are grateful to Igor Bandos, Andrei Barvinsky, Alexander Ganz, Joaquim Gomis, Purnendu Karmakar,  Mikhail Vasiliev, Peter West, Yuri Zinoviev and especially to Ruslan Metsaev and Massimo Taronna for useful discussions and comments. The authors acknowledge the warm hospitality extended to them at Lebedev Physical Institute during the final stage of this project. Work of D.S. was supported in part by the Russian Science Foundation grant 14-42-00047 in association with Lebedev Physical Institute and by the Australian Research Council project No. DP160103633.

\newpage
\section*{Appendix}
Identities involving Levi-Civita tensors
\begin{align}
&(i,j,k,l)\in\{1\,,2\}\,;\quad (a,b,c,d,e,f)\in\{0\,,1\,,2\}\nonumber\\
    &\varepsilon^{ij}\varepsilon_{ik}=\delta^j_k,\nonumber \\
    &\varepsilon^{ij}\varepsilon_{kl}=2\,\delta^i_{[k}\,\delta^j_{l]}=\delta^i_k\,\delta^j_l-\delta^i_l\,\delta^j_k, \nonumber\\
    &\varepsilon^{abc}\varepsilon_{abc}=-\,3\,! ,\nonumber\\
   &\varepsilon^{abc}\varepsilon_{abd}=-\,2\,\delta^c_d,\nonumber\\
    &\varepsilon^{abc}\varepsilon_{def}=-\,3\,!\,\delta^a_{[d}\,\delta^b_{e}\,\delta^c_{f]}
    \nonumber
\end{align}
{Charge conjugation matrix identities and rules for raising-lowering spinor indices}
\begin{align}
    &C^{-1}_{\alpha\beta}=C_{\beta\alpha} =-C_{\alpha\beta},\nonumber\\
    &\chi_\alpha=C_{\alpha \beta}\,\chi^\beta\,, \qquad  \chi^\beta=C^{\alpha \beta}\,\chi_\alpha.\nonumber
\end{align}
{$\Gamma$-matrix identities}
\begin{align}
    &\{\Gamma^a\,,\Gamma^b\}=2\,\eta^{ab}, \nonumber\\
    &\Gamma^a\,\Gamma^b\,\Gamma^c=\varepsilon^{abc}+\eta^{ab}\,\Gamma^c+\eta^{bc}\,\Gamma^a-\eta^{ac}\,\Gamma^b, \nonumber\\
    &\Gamma^a\,\Gamma^b=\varepsilon^{abc}\,\Gamma_c+\eta^{ab}, \qquad
    \varepsilon_{abc}\,\Gamma^a\,\Gamma^b=-\,2\,\Gamma_c.
    \nonumber
\end{align}

\noindent
The determinant of a $3\times 3$ matrix $E_m^a=\delta_m^a+M_m^a$
\bea\label{det3}
\det E&=&\det(1+M)=1+\tr M+\frac 12\left[(\tr M)^2-\tr (M^2)\right]\nonumber\\
&&+\frac 16\left[ (\tr M)^3 -3\tr M\,\tr(M^2)+2 \tr (M^3) \right ]
\eea


\begin{thebibliography}{10}

\bibitem{Hietarinta:1975fu}
J.~Hietarinta, \emph{{Supersymmetry Generators of Arbitrary Spin}},
\href{http://dx.doi.org/10.1103/PhysRevD.13.838}{Phys. Rev. {\bf D13} (1976)
  838}.

\bibitem{Volkov:1972jx}
D.~V. Volkov and V.~P. Akulov, \emph{{Possible universal neutrino
  interaction}},
JETP Lett. {\bf 16} (1972)  438--440.

\bibitem{Volkov:1973ix}
D.~V. Volkov and V.~P. Akulov, \emph{{Is the Neutrino a Goldstone Particle?}},
\href{http://dx.doi.org/10.1016/0370-2693(73)90490-5}{Phys. Lett. {\bf B46}
  (1973)  109--110}.

\bibitem{Baaklini:1977ka}
N.~S. Baaklini, \emph{{New Superalgebra and Goldstone Spin 3/2 Particle}},
\href{http://dx.doi.org/10.1016/0370-2693(77)90386-0}{Phys. Lett. {\bf 67B}
  (1977)  335--336}.

\bibitem{Pilot:1988st}
C.~Pilot and S.~Rajpoot, \emph{{Supersymmetry with vector-spinor generators}},
\href{http://dx.doi.org/10.1142/S021773238900037X}{Mod. Phys. Lett. {\bf A4}
  (1989)  303}.

\bibitem{Pilot:1988gy}
C.~Pilot and S.~Rajpoot, \emph{{Superalgebras with fermionic generators other
  than spin 1/2}},
\href{http://dx.doi.org/10.1142/S0217732389000976}{Mod. Phys. Lett. {\bf A4}
  (1989)  831}.

\bibitem{Shima:2000fs}
K.~Shima and M.~Tsuda, \emph{{On gravitational interaction of spin 3 / 2
  Nambu-Goldstone fermion}},
  \href{http://dx.doi.org/10.1016/S0370-2693(01)01199-6}{Phys. Lett. {\bf B521}
  (2001)  67--70},
\href{http://arxiv.org/abs/hep-th/0012235}{{\tt arXiv:hep-th/0012235
  [hep-th]}}.

\bibitem{Berends:1979wu}
F.~A. Berends, J.~W. van Holten, P.~van Nieuwenhuizen, and B.~de~Wit, \emph{{On
  spin 5/2 gauge fields}},
  \href{http://dx.doi.org/10.1016/0370-2693(79)90682-8,
  10.1016/0370-2693(79)91257-7}{Phys. Lett. {\bf 83B} (1979)  188}.
[Erratum: Phys. Lett.84B,529(1979)].

\bibitem{Berends:1979rv}
F.~A. Berends, J.~W. van Holten, P.~van Nieuwenhuizen, and B.~de~Wit, \emph{{On
  Field Theory for Massive and Massless Spin 5/2 Particles}},
\href{http://dx.doi.org/10.1016/0550-3213(79)90514-5}{Nucl. Phys. {\bf B154}
  (1979)  261--282}.

\bibitem{Berends:1979kg}
F.~A. Berends, J.~W. van Holten, B.~de~Wit, and P.~van Nieuwenhuizen, \emph{{ON
  SPIN 5/2 GAUGE FIELDS}},
\href{http://dx.doi.org/10.1088/0305-4470/13/5/022}{J. Phys. {\bf A13} (1980)
  1643--1649}.

\bibitem{Aragone:1979hx}
C.~Aragone and S.~Deser, \emph{{Consistency Problems of Hypergravity}},
\href{http://dx.doi.org/10.1016/0370-2693(79)90808-6}{Phys. Lett. {\bf B86}
  (1979)  161}.

\bibitem{Aragone:1983sz}
C.~Aragone and S.~Deser, \emph{{Hypersymmetry in $D=3$ of Coupled Gravity
  Massless Spin 5/2 System}},
\href{http://dx.doi.org/10.1088/0264-9381/1/2/001}{Class. Quant. Grav. {\bf 1}
  (1984)  L9}.

\bibitem{Zinoviev:2014sza}
{\relax Yu}.~M. Zinoviev, \emph{{Hypergravity in AdS$_3$}},
  \href{http://dx.doi.org/10.1016/j.physletb.2014.10.041}{Phys. Lett. {\bf
  B739} (2014)  106--109},
\href{http://arxiv.org/abs/1408.2912}{{\tt arXiv:1408.2912 [hep-th]}}.

\bibitem{Bunster:2014fca}
C.~Bunster, M.~Henneaux, S.~Hortner, and A.~Leonard, \emph{{Supersymmetric
  electric-magnetic duality of hypergravity}},
  \href{http://dx.doi.org/10.1103/PhysRevD.95.069908,
  10.1103/PhysRevD.90.045029}{Phys. Rev. {\bf D90} (2014) no.~4, 045029},
  \href{http://arxiv.org/abs/1406.3952}{{\tt arXiv:1406.3952 [hep-th]}}.
[Erratum: Phys. Rev.D95,no.6,069908(2017)].

\bibitem{Fuentealba:2015jma}
O.~Fuentealba, J.~Matulich, and R.~Troncoso, \emph{{Extension of the Poincare
  group with half-integer spin generators: hypergravity and beyond}},
  \href{http://dx.doi.org/10.1007/JHEP09(2015)003}{JHEP {\bf 09} (2015)  003},
\href{http://arxiv.org/abs/1505.06173}{{\tt arXiv:1505.06173 [hep-th]}}.

\bibitem{Fuentealba:2015wza}
O.~Fuentealba, J.~Matulich, and R.~Troncoso, \emph{{Asymptotically flat
  structure of hypergravity in three spacetime dimensions}},
  \href{http://dx.doi.org/10.1007/JHEP10(2015)009}{JHEP {\bf 10} (2015)  009},
\href{http://arxiv.org/abs/1508.04663}{{\tt arXiv:1508.04663 [hep-th]}}.

\bibitem{Henneaux:2015tar}
M.~Henneaux, A.~P\'erez, D.~Tempo, and R.~Troncoso,
  \href{http://dx.doi.org/10.1142/9789813144101_0009}{\emph{{Extended anti-de
  Sitter Hypergravity in $2+1$ Dimensions and Hypersymmetry Bounds}},} in {\em
  {Proceedings, International Workshop on Higher Spin Gauge Theories:
  Singapore, Singapore, November 4-6, 2015}}, pp.~139--157.
\newblock 2017.
\newblock
\href{http://arxiv.org/abs/1512.08603}{{\tt arXiv:1512.08603 [hep-th]}}.
\newblock

\bibitem{Vasiliev:1995dn}
M.~A. Vasiliev, \emph{{Higher--spin gauge theories in four, three and two
  dimensions}}, \href{http://dx.doi.org/10.1142/S0218271896000473}{Int. J. Mod.
  Phys. {\bf D5} (1996)  763--797},
\href{http://arxiv.org/abs/hep-th/9611024}{{\tt arXiv:hep-th/9611024}}.

\bibitem{Buchbinder:1995uq}
I.~Buchbinder and S.~Kuzenko, \emph{{Ideas and methods of supersymmetry and
  supergravity: A Walk through superspace}},
1995  .

\bibitem{Vasiliev:1999ba}
M.~A. Vasiliev, \emph{{Higher spin gauge theories: Star product and AdS
  space}}, \href{http://arxiv.org/abs/hep-th/9910096}{{\tt arXiv:hep-th/9910096
  [hep-th]}}.
Contributed article to Golfand's Memorial Volume, M. Shifman ed., World
  Scientific.

\bibitem{Vasiliev:2001ur}
M.~A. Vasiliev, \emph{{Progress in higher spin gauge theories}},
\href{http://arxiv.org/abs/hep-th/0104246}{{\tt arXiv:hep-th/0104246}}.

\bibitem{Bekaert:2003uc}
X.~Bekaert, I.~L. Buchbinder, A.~Pashnev, and M.~Tsulaia, \emph{{On higher spin
  theory: Strings, BRST, dimensional reductions}}, Class. Quant. Grav. {\bf 21}
  (2004)  S1457--1464,
\href{http://arxiv.org/abs/hep-th/0312252}{{\tt arXiv:hep-th/0312252}}.

\bibitem{Sorokin:2004ie}
D.~Sorokin, \emph{{Introduction to the classical theory of higher spins}},
  \href{http://dx.doi.org/10.1063/1.1923335}{AIP Conf. Proc. {\bf 767} (2005)
  172--202},
\href{http://arxiv.org/abs/hep-th/0405069}{{\tt arXiv:hep-th/0405069}}.

\bibitem{Bouatta:2004kk}
N.~Bouatta, G.~Compere, and A.~Sagnotti, \emph{{An introduction to free
  higher-spin fields}},
\href{http://arxiv.org/abs/hep-th/0409068}{{\tt arXiv:hep-th/0409068}}.

\bibitem{Bandos:2005rr}
I.~A. Bandos, \emph{{BPS preons in supergravity and higher spin theories. An
  Overview from the hill of twistor appraoch}},
  \href{http://dx.doi.org/10.1063/1.1923334}{AIP Conf. Proc. {\bf 767} (2005)
  141--171}, \href{http://arxiv.org/abs/hep-th/0501115}{{\tt
  arXiv:hep-th/0501115 [hep-th]}}.
[,141(2005)].

\bibitem{Bekaert:2005vh}
X.~Bekaert, S.~Cnockaert, C.~Iazeolla, and M.~A. Vasiliev, \emph{{Nonlinear
  higher spin theories in various dimensions}},
\href{http://arxiv.org/abs/hep-th/0503128}{{\tt arXiv:hep-th/0503128}}.

\bibitem{Francia:2006hp}
D.~Francia and A.~Sagnotti, \emph{{Higher-spin geometry and string theory}},
  \href{http://dx.doi.org/10.1088/1742-6596/33/1/006}{J. Phys. Conf. Ser. {\bf
  33} (2006)  57},
\href{http://arxiv.org/abs/hep-th/0601199}{{\tt arXiv:hep-th/0601199
  [hep-th]}}.

\bibitem{Fotopoulos:2008ka}
A.~Fotopoulos and M.~Tsulaia, \emph{{Gauge Invariant Lagrangians for Free and
  Interacting Higher Spin Fields. A Review of the BRST formulation}},
  \href{http://dx.doi.org/10.1142/S0217751X09043134}{Int. J. Mod. Phys. {\bf
  A24} (2009)  1--60},
\href{http://arxiv.org/abs/0805.1346}{{\tt arXiv:0805.1346 [hep-th]}}.

\bibitem{Campoleoni:2009je}
A.~Campoleoni, \emph{{Metric-like Lagrangian Formulations for Higher-Spin
  Fields of Mixed Symmetry}},
  \href{http://dx.doi.org/10.1393/ncr/i2010-10053-2}{Riv.Nuovo Cim. {\bf 033}
  (2010)  123--253}, \href{http://arxiv.org/abs/0910.3155}{{\tt arXiv:0910.3155
  [hep-th]}}.

\bibitem{Taronna:2010qq}
M.~Taronna, \emph{{Higher Spins and String Interactions}},
\href{http://arxiv.org/abs/1005.3061}{{\tt arXiv:1005.3061 [hep-th]}}.

\bibitem{Bekaert:2010hw}
X.~Bekaert, N.~Boulanger, and P.~Sundell, \emph{{How higher-spin gravity
  surpasses the spin two barrier: no-go theorems versus yes-go examples}},
  \href{http://dx.doi.org/10.1103/RevModPhys.84.987}{Rev.Mod.Phys. {\bf 84}
  (2012)  987--1009},
\href{http://arxiv.org/abs/1007.0435}{{\tt arXiv:1007.0435 [hep-th]}}.

\bibitem{Sezgin:2012ag}
E.~Sezgin and P.~Sundell, \emph{{Supersymmetric Higher Spin Theories}},
  \href{http://dx.doi.org/10.1088/1751-8113/46/21/214022}{J.Phys. {\bf A46}
  (2013)  214022},
\href{http://arxiv.org/abs/1208.6019}{{\tt arXiv:1208.6019 [hep-th]}}.

\bibitem{Gaberdiel:2012uj}
M.~R. Gaberdiel and R.~Gopakumar, \emph{{Minimal Model Holography}},
  \href{http://dx.doi.org/10.1088/1751-8113/46/21/214002}{J. Phys. {\bf A46}
  (2013)  214002},
\href{http://arxiv.org/abs/1207.6697}{{\tt arXiv:1207.6697 [hep-th]}}.

\bibitem{Giombi:2012ms}
S.~Giombi and X.~Yin, \emph{{The Higher Spin/Vector Model Duality}},
  \href{http://dx.doi.org/10.1088/1751-8113/46/21/214003}{J. Phys. {\bf A46}
  (2013)  214003},
\href{http://arxiv.org/abs/1208.4036}{{\tt arXiv:1208.4036 [hep-th]}}.

\bibitem{Joung:2012fv}
E.~Joung, L.~Lopez, and M.~Taronna, \emph{{Solving the Noether procedure for
  cubic interactions of higher spins in (A)dS}},
  \href{http://dx.doi.org/10.1088/1751-8113/46/21/214020}{J. Phys. {\bf A46}
  (2013)  214020},
\href{http://arxiv.org/abs/1207.5520}{{\tt arXiv:1207.5520 [hep-th]}}.

\bibitem{Sagnotti:2013bha}
A.~Sagnotti, \emph{{Notes on Strings and Higher Spins}},
  \href{http://dx.doi.org/10.1142/9789814522519_0008,
  10.1088/1751-8113/46/21/214006}{J. Phys. {\bf A46} (2013)  214006},
\href{http://arxiv.org/abs/1112.4285}{{\tt arXiv:1112.4285 [hep-th]}}.

\bibitem{Didenko:2014dwa}
V.~E. Didenko and E.~D. Skvortsov, \emph{{Elements of Vasiliev theory}},
\href{http://arxiv.org/abs/1401.2975}{{\tt arXiv:1401.2975 [hep-th]}}.

\bibitem{Rahman:2015pzl}
R.~Rahman and M.~Taronna, \emph{{From Higher Spins to Strings: A Primer}},
\href{http://arxiv.org/abs/1512.07932}{{\tt arXiv:1512.07932 [hep-th]}}.

\bibitem{Arias:2016ajh}
C.~Arias, R.~Bonezzi, N.~Boulanger, E.~Sezgin, P.~Sundell, A.~Torres-Gomez, and
  M.~Valenzuela,
  \href{http://dx.doi.org/10.1142/9789813144101_0012}{\emph{{Action principles
  for higher and fractional spin gravities}},} in {\em {Proceedings,
  International Workshop on Higher Spin Gauge Theories: Singapore, Singapore,
  November 4-6, 2015}}, pp.~213--253.
\newblock 2017.
\newblock
\href{http://arxiv.org/abs/1603.04454}{{\tt arXiv:1603.04454 [hep-th]}}.
\newblock

\bibitem{Sleight:2016hyl}
C.~Sleight, \emph{{Interactions in Higher-Spin Gravity: a Holographic
  Perspective}}, \href{http://dx.doi.org/10.1088/1751-8121/aa820c}{J. Phys.
  {\bf A50} (2017) no.~38, 383001},
\href{http://arxiv.org/abs/1610.01318}{{\tt arXiv:1610.01318 [hep-th]}}.

\bibitem{Sleight:2017krf}
C.~Sleight, \emph{{Metric-like Methods in Higher Spin Holography}}, PoS {\bf
  Modave2016} (2017)  003,
\href{http://arxiv.org/abs/1701.08360}{{\tt arXiv:1701.08360 [hep-th]}}.

\bibitem{Sorokin:2017irs}
D.~Sorokin and M.~Tsulaia, \emph{{Higher Spin Fields in Hyperspace. A Review}},
  \href{http://dx.doi.org/10.3390/universe4010007}{Universe {\bf 4} (2018)
  no.~1, 7},
\href{http://arxiv.org/abs/1710.08244}{{\tt arXiv:1710.08244 [hep-th]}}.

\bibitem{Rahman:2017cxk}
R.~Rahman, \emph{{Frame- and Metric-like Higher-Spin Fermions}},
\href{http://arxiv.org/abs/1712.09264}{{\tt arXiv:1712.09264 [hep-th]}}.

\bibitem{Shima:1989nb}
K.~Shima and Y.~Tanii, \emph{{Unitarity Constraint on General Supersymmetry
  Algebra}},
\href{http://dx.doi.org/10.1142/S0217732389002549}{Mod. Phys. Lett. {\bf A4}
  (1989)  2259}.

\bibitem{Dirac-lectures}
P.~Dirac, {\em Lectures on Quantum Mechanics}.
\newblock Dover Publications Inc. Mineola, New York, 2001.

\bibitem{HT-book}
M.~Henneaux and C.~Teitelboim, {\em Quantization of Gauge Systems}.
\newblock Princeton University Press, 1994.

\bibitem{Nicolis:2008in}
A.~Nicolis, R.~Rattazzi, and E.~Trincherini, \emph{{The Galileon as a local
  modification of gravity}},
  \href{http://dx.doi.org/10.1103/PhysRevD.79.064036}{Phys. Rev. {\bf D79}
  (2009)  064036},
\href{http://arxiv.org/abs/0811.2197}{{\tt arXiv:0811.2197 [hep-th]}}.

\bibitem{Hinterbichler:2011tt}
K.~Hinterbichler, \emph{{Theoretical Aspects of Massive Gravity}},
  \href{http://dx.doi.org/10.1103/RevModPhys.84.671}{Rev. Mod. Phys. {\bf 84}
  (2012)  671--710},
\href{http://arxiv.org/abs/1105.3735}{{\tt arXiv:1105.3735 [hep-th]}}.

\bibitem{deRham:2014zqa}
C.~de~Rham, \emph{{Massive Gravity}},
  \href{http://dx.doi.org/10.12942/lrr-2014-7}{Living Rev. Rel. {\bf 17} (2014)
   7},
\href{http://arxiv.org/abs/1401.4173}{{\tt arXiv:1401.4173 [hep-th]}}.

\bibitem{Schmidt-May:2015vnx}
A.~Schmidt-May and M.~von Strauss, \emph{{Recent developments in bimetric
  theory}}, \href{http://dx.doi.org/10.1088/1751-8113/49/18/183001}{J. Phys.
  {\bf A49} (2016) no.~18, 183001},
\href{http://arxiv.org/abs/1512.00021}{{\tt arXiv:1512.00021 [hep-th]}}.

\bibitem{Amendola:2016saw}
L.~Amendola {\em et al.}, \emph{{Cosmology and fundamental physics with the
  Euclid satellite}}, \href{http://dx.doi.org/10.1007/s41114-017-0010-3}{Living
  Rev. Rel. {\bf 21} (2018) no.~1, 2},
\href{http://arxiv.org/abs/1606.00180}{{\tt arXiv:1606.00180 [astro-ph.CO]}}.

\bibitem{Bandos:2015xnf}
I.~Bandos, L.~Martucci, D.~Sorokin, and M.~Tonin, \emph{{Brane induced
  supersymmetry breaking and de Sitter supergravity}},
  \href{http://dx.doi.org/10.1007/JHEP02(2016)080}{JHEP {\bf 02} (2016)  080},
\href{http://arxiv.org/abs/1511.03024}{{\tt arXiv:1511.03024 [hep-th]}}.

\bibitem{Bandos:2016xyu}
I.~Bandos, M.~Heller, S.~M. Kuzenko, L.~Martucci, and D.~Sorokin, \emph{{The
  Goldstino brane, the constrained superfields and matter in $ \mathcal{N}=1 $
  supergravity}}, \href{http://dx.doi.org/10.1007/JHEP11(2016)109}{JHEP {\bf
  11} (2016)  109},
\href{http://arxiv.org/abs/1608.05908}{{\tt arXiv:1608.05908 [hep-th]}}.

\bibitem{Horndeski:1974wa}
G.~W. Horndeski, \emph{{Second-order scalar-tensor field equations in a
  four-dimensional space}},
\href{http://dx.doi.org/10.1007/BF01807638}{Int. J. Theor. Phys. {\bf 10}
  (1974)  363--384}.

\bibitem{Gleyzes:2014dya}
J.~Gleyzes, D.~Langlois, F.~Piazza, and F.~Vernizzi, \emph{{Healthy theories
  beyond Horndeski}},
  \href{http://dx.doi.org/10.1103/PhysRevLett.114.211101}{Phys. Rev. Lett. {\bf
  114} (2015) no.~21, 211101},
\href{http://arxiv.org/abs/1404.6495}{{\tt arXiv:1404.6495 [hep-th]}}.

\bibitem{Zhou:2010di}
S.-Y. Zhou, \emph{{Goldstone's Theorem and Hamiltonian of Multi-galileon
  Modified Gravity}}, \href{http://dx.doi.org/10.1103/PhysRevD.83.064005}{Phys.
  Rev. {\bf D83} (2011)  064005},
\href{http://arxiv.org/abs/1011.0863}{{\tt arXiv:1011.0863 [hep-th]}}.

\bibitem{Sivanesan:2011kw}
V.~Sivanesan, \emph{{Hamiltonian of galileon field theory}},
  \href{http://dx.doi.org/10.1103/PhysRevD.85.084018}{Phys. Rev. {\bf D85}
  (2012)  084018},
\href{http://arxiv.org/abs/1111.3558}{{\tt arXiv:1111.3558 [hep-th]}}.

\bibitem{Achucarro:1987vz}
A.~Achucarro and P.~K. Townsend, \emph{{A Chern-Simons Action for
  Three-Dimensional anti-De Sitter Supergravity Theories}},
  \href{http://dx.doi.org/10.1016/0370-2693(86)90140-1}{Phys. Lett. {\bf B180}
  (1986)  89}.
[732(1987)].

\bibitem{Witten:1988hc}
E.~Witten, \emph{{(2+1)-Dimensional Gravity as an Exactly Soluble System}},
\href{http://dx.doi.org/10.1016/0550-3213(88)90143-5}{Nucl. Phys. {\bf B311}
  (1988)  46}.

\bibitem{Bergshoeff:2009hq}
E.~A. Bergshoeff, O.~Hohm, and P.~K. Townsend, \emph{{Massive Gravity in Three
  Dimensions}}, \href{http://dx.doi.org/10.1103/PhysRevLett.102.201301}{Phys.
  Rev. Lett. {\bf 102} (2009)  201301},
\href{http://arxiv.org/abs/0901.1766}{{\tt arXiv:0901.1766 [hep-th]}}.

\bibitem{Bergshoeff:2009aq}
E.~A. Bergshoeff, O.~Hohm, and P.~K. Townsend, \emph{{More on Massive 3D
  Gravity}}, \href{http://dx.doi.org/10.1103/PhysRevD.79.124042}{Phys. Rev.
  {\bf D79} (2009)  124042},
\href{http://arxiv.org/abs/0905.1259}{{\tt arXiv:0905.1259 [hep-th]}}.

\bibitem{Bergshoeff:2013xma}
E.~A. Bergshoeff, S.~de~Haan, O.~Hohm, W.~Merbis, and P.~K. Townsend,
  \emph{{Zwei-Dreibein Gravity: A Two-Frame-Field Model of 3D Massive
  Gravity}}, \href{http://dx.doi.org/10.1103/PhysRevLett.111.111102,
  10.1103/PhysRevLett.111.259902}{Phys. Rev. Lett. {\bf 111} (2013) no.~11,
  111102}, \href{http://arxiv.org/abs/1307.2774}{{\tt arXiv:1307.2774
  [hep-th]}}.
[Erratum: Phys. Rev. Lett.111,no.25,259902(2013)].

\bibitem{Ozkan:2018cxj}
M.\"Ozkan, Y.~Pang, and P.~K. Townsend, \emph{{Exotic Massive 3D Gravity}},
\href{http://arxiv.org/abs/1806.04179}{{\tt arXiv:1806.04179 [hep-th]}}.

\bibitem{Ivanov:1984nu}
E.~A. Ivanov and J.~Niederle, \emph{{Construction of the superalgebras for N=1
  supergravity}},
\href{http://dx.doi.org/10.1088/0264-9381/2/5/006}{Class. Quant. Grav. {\bf 2}
  (1985)  631}.

\bibitem{Ivanov:1992ax}
E.~A. Ivanov and J.~Niederle, \emph{{N=1 supergravity as a nonlinear
  realization}},
\href{http://dx.doi.org/10.1103/PhysRevD.45.4545}{Phys. Rev. {\bf D45} (1992)
  4545--4554}.

\bibitem{Maxera:1994xv}
D.~Maxera and J.~Niederle, \emph{{N=1 supergravity as a nonlinear realization.
  2. The General case}},
\href{http://dx.doi.org/10.1103/PhysRevD.50.6318}{Phys. Rev. {\bf D50} (1994)
  6318--6328}.

\end{thebibliography}
\providecommand{\href}[2]{#2}\begingroup\raggedright\endgroup

\end{document}